# uTECH Turbo-MOVES: The Intersection of Software Architectural Innovation and Sustainable Mobility Analysis

Cafer Avcı[a], Alireza Yazdiani[b], Roopak Srinivasan[b, c], Megh Khaire[b, c], Mohammad Tayarani[b], H. Oliver Gao[b, d, *]

[a] Systems Engineering, Cornell University, Ithaca, NY, 14853, USA
[b] School of Civil and Environmental Engineering, Cornell University, Ithaca, NY, 14853, USA
[c] College of Computing and Information Science, Ithaca, NY, 14853, USA
[d] Center for Transportation, Environment, and Community Health (CTECH), Cornell University, 313 Hollister Hall, Ithaca, NY, 14853-3801, USA

**Abstract**

Transportation is a major source of energy consumption, greenhouse gas emissions, and criteria air pollutants, making efficient emissions estimation essential for environmental planning and policy analyses. The U.S. Environmental Protection Agency's Motor Vehicle Emission Simulator (MOVES) supports detailed emissions modeling, but its computational demands constrain large-scale scenario analyses. This study introduces uTECH Turbo-MOVES, a software architecture that expedites MOVES computation without modifying its Calculator modules or input/output structure. By combining RunSpec partitioning, container isolation, in-memory intermediate storage, and concurrent workload orchestration, it improves execution efficiency. We evaluate six benchmark cases spanning County, Default, and Project scales and both Emission Inventory and Emission Rates modes. Results demonstrate a measured 5.5× and an estimated 15.3× per-run speedup for Default-scale workloads, and an 84.8× campaign speedup for an eight-county Emission Rates analysis. Although gains vary by workload and hardware configuration, uTECH Turbo-MOVES enables more extensive environmental assessments and transportation policy evaluation.

* Corresponding Author: hg55@cornell.edu

## 1. Introduction

Transportation is the largest source of greenhouse gas (GHG) emissions in the United States, accounting for 28% of national emissions in 2022 (US EPA, 2025a), and it remains a major contributor to urban air pollution, with adverse public-health effects that fall disproportionately on marginalized communities (Khan et al., 2024; Tayarani et al., 2016). Mitigating these externalities requires coordinated action at the federal, state, and local levels (Tayarani et al., 2018). Supply-side standards such as the Corporate Average Fuel Economy program and California's Advanced Clean Cars regulations constrain vehicle emissions (Bui and Yang, 2022; NHTSA, 2025); demand-side measures under the Inflation Reduction Act incentivize electric-vehicle adoption (Allcott et al., 2024; StandfordReport, 2024); and local instruments, including congestion pricing and investment in transit and active transportation, manage travel demand in major urban areas (Cook et al., 2025; Gu et al., 2018; Winkless, 2025). The adoption of such measures depends in part on credible evidence of their emission consequences.

Producing that evidence has become a major analytical bottleneck. Emissions modeling for urban transportation systems, particularly in large metropolitan areas, is time- and resource-intensive (Liu et al., 2017), and practitioners consequently limit their analyses to a small number of scenarios, constrained by methodological complexity, computational burden, and project timelines; as a result, many questions raised by stakeholders remain unanswered (Nadafianshahamabadi, 2019). This challenge will intensify as connected and autonomous vehicles, shared mobility, and electrification increase the complexity of transportation systems and push practice toward digital-twin applications that require near-interactive model turnaround (Wu et al., 2025). In the United States, such analyses center on EPA MOVES, the U.S. Environmental Protection Agency's Motor Vehicle Emission Simulator, whose use is required for regulatory applications including State Implementation Plan inventories and transportation conformity determinations. The computational cost of MOVES, rather than any shortage of policy questions, is therefore the binding constraint on scenario analysis, and it has motivated a decade of acceleration research.

Prior acceleration efforts fall into three classes. The first employs high-performance computing (HPC) to pre-compute MOVES outputs exhaustively. MOVES-Matrix pre-executed every combination of vehicle class, fuel type, meteorological condition, and operating mode on a supercomputing cluster, producing multidimensional lookup tables that return link- and inventory-level emission rates approximately 200 times faster than the MOVES graphical interface, and the same exhaustive strategy has been applied to build region-specific emission-factor repositories, achieving large reductions in conformity-analysis runtimes at the price of storage requirements exceeding 100 GB and a continuing dependence on HPC access (Liu et al., 2019; Xu et al., 2016).

The second class replaces portions of MOVES's internal logic with simplified analytical formulations. MOVEStar, for example, is a lightweight re-implementation that fixes several default parameters (e.g., region, fuel, and road grade) and operates on second-by-second speed traces and vehicle classifications (Wang et al., 2021); the design sacrifices configurability but achieves quasi-real-time performance for project-level analysis on standard desktop hardware.

The third class, and the most active research frontier, comprises statistical and machine-learning surrogates. Zhong et al. (2024) demonstrated that link-level random-forest and generalized-linear models trained on hourly traffic descriptors (average speed, speed variability, heavy-duty share, and road grade) can replicate MOVES inventories with planning-grade fidelity. MOVESLite fits county- and pollutant-specific ensembles of regularized linear models to a large corpus of default MOVES runs, reducing each county-year estimate to approximately one millisecond and enabling high-throughput policy screening (Fraser et al., 2024). Zhong et al. (2024) further reduced computational demand by clustering roadway links with homogeneous emission signatures and applying MOVES only to representative members of each cluster, and NeuralMOVES distills more than two million reverse-engineered scenarios into a 2.4 MB neural network that reproduces $CO_2$ rates within 6% of MOVES and supports second-by-second control applications (Ramirez-Sanchez et al., 2025).

Each class reduces MOVES runtimes from hours to seconds, and each introduces a distinct trade-off: extensive storage and specialized hardware for HPC lookup tables, reduced input flexibility for streamlined re-implementations, and large training datasets with meticulous validation for surrogate models. More fundamentally, all three approximate or replace MOVES, so their outputs cannot substitute for the official MOVES model where exact regulatory fidelity is required; surrogates, moreover, themselves depend on collections of default MOVES runs for training and periodic recalibration. Selecting among them therefore requires weighing analytical scope, available computational resources, and the acceptable deviation from the default model. What has been missing is an approach that accelerates MOVES itself.

To fill this gap, we introduce uTECH Turbo-MOVES, an optimized computing architecture developed within the urban Transportation, Environment, and Community Health (uTECH) modeling initiative. Rather than approximating the model, uTECH Turbo-MOVES pursues what we term architectural acceleration by modifying how a MOVES run is partitioned, deployed, and stored (through RunSpec splitting, a native Linux port, containerized execution, in-memory storage, and parallel cloud orchestration) while leaving MOVES's Calculator modules and input/output structure unmodified, so that its outputs are identical to those of the MOVES model by construction and its regulatory use is unaffected. Because it preserves the MOVES model exactly, the architecture complements rather than competes with the approaches above. It retains full regulatory validity, it can generate the large collections of MOVES runs on which surrogates such as MOVESLite are trained and recalibrated, and it provides a computational foundation for the digital-twin applications toward which practice is moving.

This study presents the design of uTECH Turbo-MOVES and evaluates its computational performance across six combinations of MOVES analysis scale (Default, County, and Project) and calculation type (Inventory and Emission Rates). The evaluation uses benchmark scenarios for an eight-county New York metropolitan region and a highway interchange application. Section 2 describes the MOVES computational architecture and its bottlenecks, introduces the uTECH Turbo-MOVES design, and details the experimental methods. Section 3 presents the computational performance results. Section 4 interprets the findings and discusses their practical implications. Section 5 summarizes the conclusions and future research directions.

## 2. Methods

This section describes the software design of uTECH Turbo-MOVES and the experimental design used to evaluate it. Section 2.1 summarizes the EPA MOVES model and its execution architecture, and Section 2.2 examines the computational limitations of that architecture. Section 2.3 presents the architectural modifications that constitute uTECH Turbo-MOVES, and Section 2.4 describes the experimental design.

## 2.1. The EPA MOVES model and its computation architecture

MOVES is the U.S. Environmental Protection Agency's official model for estimating emissions and energy consumption from on-road mobile sources and is used for regulatory applications, including State Implementation Plan inventories and transportation conformity analyses. MOVES6, initially released in September 2026, is the latest version, succeeding MOVES5, which was released in November 2024 (US EPA, 2025b). The model estimates emissions by combining emission factors with vehicle activity data. Emission factors are organized by vehicle type, vehicle age, fuel type, operating mode, and road type, and are further adjusted for vehicle speed and acceleration, ambient temperature and humidity, fuel properties, and emission control technologies (Fujita et al., 2012; Hall-Quinlan et al., 2023; Zhao and Sadek, 2013). The comprehensiveness of these calculations makes MOVES applications for large urban areas time-consuming and resource-intensive (Liu et al., 2019, 2017).

From a computational perspective, MOVES is a Java application built on a client–server architecture with a relational (MySQL/MariaDB) database backend (Fig. 1). A model run is specified through a RunSpec, an XML file that defines the geographic domain, time span, vehicle and fuel coverage, pollutants, and output detail, and that can be supplied through a graphical user interface or the command line. The MOVES Main program decomposes a run into work bundles, which one or more MOVES Worker processes execute using the model's Calculator modules and return as completed bundles; multiple Workers sharing a common work directory can therefore process bundles in parallel. The model uses databases throughout execution. The Main program draws on the Default database, which contains national inputs for calendar years 1990 and 1999–2060, together with a user-supplied domain database. Intermediate state is staged in a temporary Execution database, and each Worker maintains its own working database. Results are written to a user-named Output database, in the movesoutput table for Inventory mode runs or in a set of rate tables for Emission Rates mode runs.

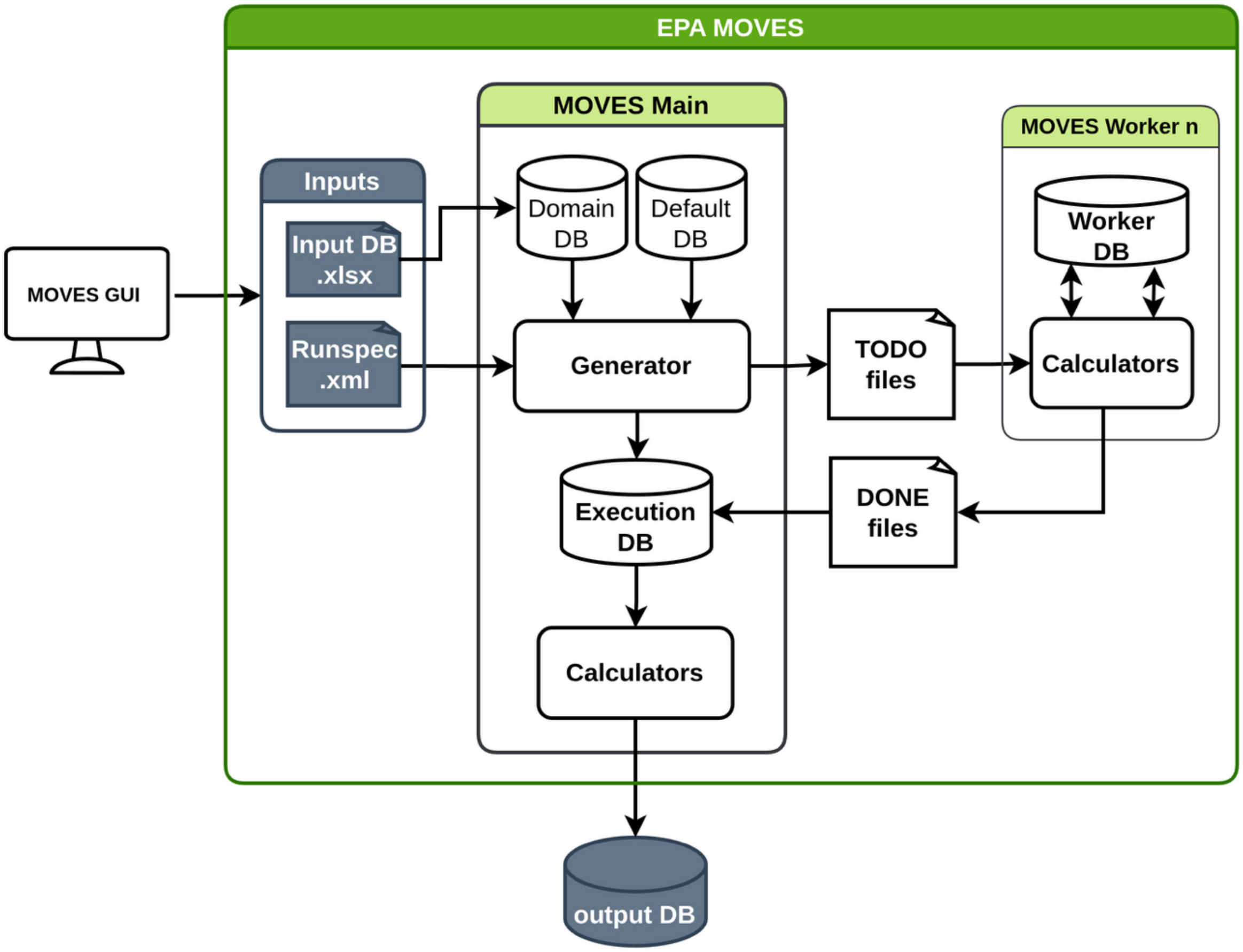

**Figure 1.** Computation architecture of the default EPA MOVES distribution.

## 2.2. Computational bottlenecks in the default architecture

The performance of this architecture becomes limiting at the scales required by regulatory practice. A regional conformity analysis for a metropolitan planning organization can require days to weeks of computation, and the inclusion of additional pollutants, months, or day types compounds this burden. Three properties of the design explain these limitations.

First, the internal parallelism of MOVES saturates at a small number of Worker processes. To quantify this ceiling, we conducted a worker-scaling experiment for a Bronx County run on a high-performance workstation with 52 CPU cores and 250 GB of memory, varying the number of Worker processes attached to a single Main instance (Table 1). Increasing the number of workers from one to three reduced total runtime by 17.4%, from 178 to 147 min. Using more than three workers increased runtime relative to the three-worker configuration. Because all Workers operate against shared database resources, higher concurrency produces lock contention and synchronization overhead rather than additional throughput. The default architecture is therefore unable to exploit more than a small fraction of the processor cores available to it, regardless of the underlying hardware. All baseline configurations in this study accordingly use the empirically optimal configuration of one Main and three Worker processes.

**Table 1** Total runtime of a Bronx County EPA MOVES run as a function of Worker count on a 52-core workstation.

| Number of Workers | Total runtime (min) |
|---|---|
| 1 | 178 |
| 2 | 151 |
| 3 | 147 |
| 4 | 153 |
| 5 | 168 |
| 6 | 169 |

Second, MOVES is not designed to execute multiple Main instances concurrently on a single machine. Because the Main process is not sandboxed, critical resources such as the execution database, temporary directories, and caches would be shared between concurrent runs; to prevent such interference, MOVES binds port 13131 so that only one Main process can execute at a time (US EPA, 2025b). Multi-county and multi-scenario studies are therefore forced to execute serially on any single machine.

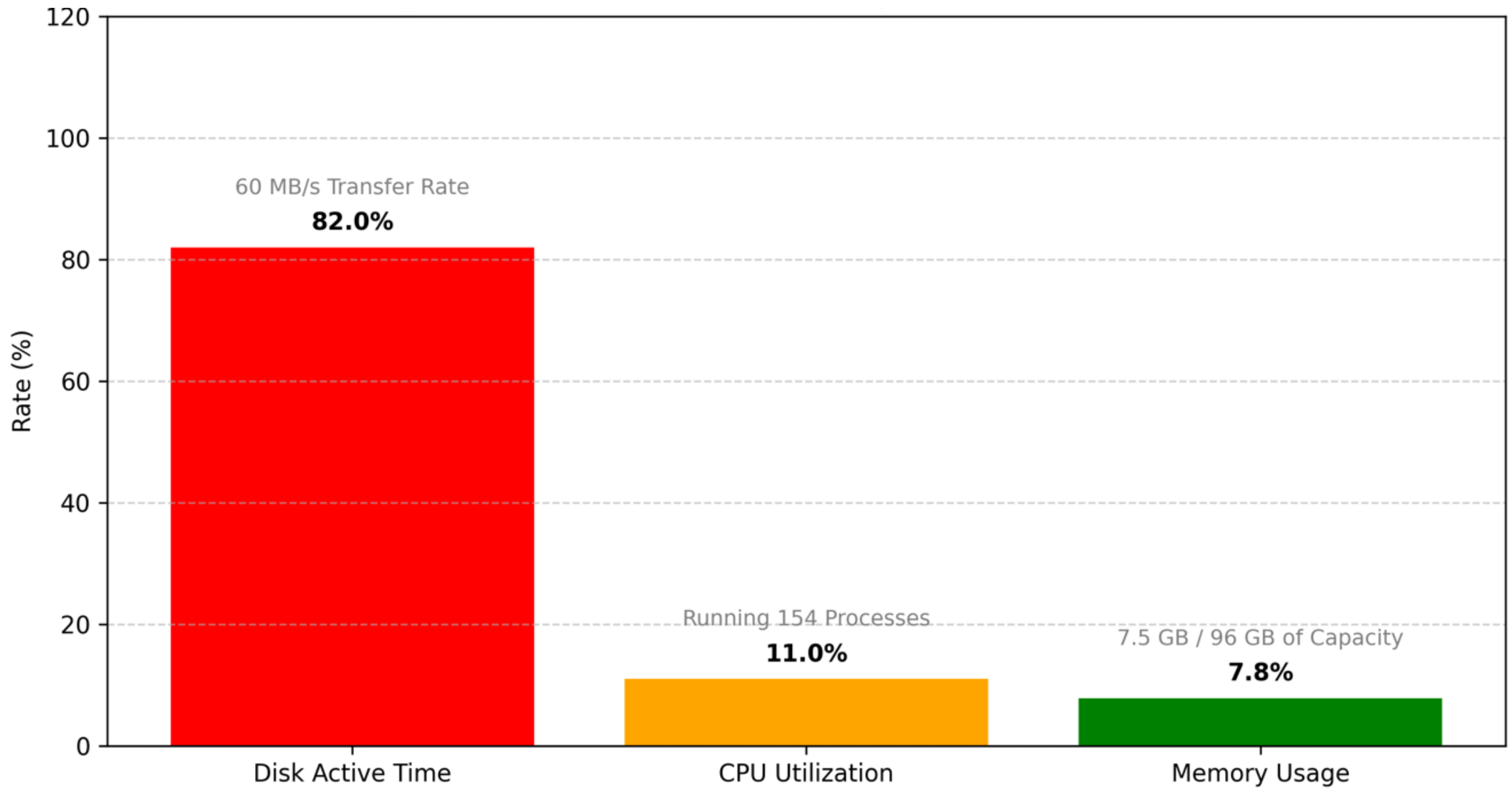

**Figure 2.** System resource utilization during a county-level EPA MOVES run.

Third, execution is bound by input/output (I/O) rather than by computation. Because intermediate results flow continuously through disk-resident database tables, a MOVES run issues hundreds of small disk operations per second. Profiling of a representative county-level run (Fig. 2) shows disk-active time averaging 82% with periodic spikes approaching 100%, whereas CPU utilization peaks near 11% and memory use remains minimal at 7.6 GB of the 96 GB available. The processors are predominantly idle while waiting on storage, so the limiting resource is the storage subsystem rather than processing capacity or memory.

Together these properties explain why practitioners routinely restrict the number of scenarios they analyze, and they motivate an acceleration strategy that modifies how MOVES is deployed without altering what it computes.

## 2.3. The uTECH Turbo-MOVES architecture

Unlike the approximation and replacement approaches reviewed in Section 1 (Fraser et al., 2024; Liu et al., 2019; Ramirez-Sanchez et al., 2025; Wang et al., 2021), uTECH Turbo-MOVES pursues architectural acceleration; it modifies how a model run is partitioned, deployed, and stored, while leaving the Calculator modules unmodified, so that its outputs are identical to those of the default model by construction. The architecture comprises four cumulative modifications, described in Sections 2.3.1 to 2.3.4 and integrated into a single workflow in Section 2.3.5.

### 2.3.1 RunSpec splitting for parallelization and memory efficiency

To address memory constraints and improve parallel execution efficiency, a single large RunSpec is partitioned into smaller, independent sub-RunSpecs along the source-type dimension, the pollutant dimension, or both. Partitioning along the pollutant dimension respects the pollutant dependency chains that MOVES enforces, so that carbon dioxide equivalent, for example, requires carbon dioxide, methane, and nitrous oxide to remain within the same sub-RunSpec and each sub-run remains self-contained.

Splitting mitigates a second source of inefficiency in addition to enabling distribution across compute nodes. MOVES performs many of its core calculations through relational database operations, including large joints. Join performance is sensitive to the size of the input and intermediate relations and to the memory available for processing; when working data no longer fit efficiently in memory, database systems may incur substantial additional I/O from partitioning or spilling intermediate state to storage (Barber et al., 2014). Smaller sub-runs reduce the size of intermediate tables, allowing more of the working data to remain in memory and reducing the I/O burden associated with large database operations. As a result, work bundles can be generated more quickly by the Main program and processed more efficiently by the Workers while preserving the same final outputs.

### 2.3.2 Linux compatibility for efficient cloud deployment

MOVES targets the Windows operating system primarily to support its graphical user interface, which automated, large-scale simulation workflows do not require. We therefore ported MOVES to run natively on Linux, removing this dependency and enabling the lightweight resource management, fast provisioning, and mature containerization ecosystem of Linux-based cloud infrastructure. Across the benchmark cases reported in Section 3, changing the operating environment reduced runtime by 14–31% (a 1.2–1.5× speedup) in Cases A, B, D, and E, whereas Case C ran slightly slower on Linux (Table 4).

### 2.3.3 Containerization for scalability and isolation

With RunSpec splitting in place and MOVES ported to Linux, each sub-run executes in its own Docker container with a private MOVES instance, database service, temporary directories, and caches. This sandboxing eliminates the shared-resource conflicts that the port-13131 restriction is designed to guard against, allowing many MOVES instances to run concurrently on a single machine and substantially improving CPU utilization across processor cores. Containerized runs nevertheless remained I/O-bound, because the volume of disk operations scales with the number of concurrent containers. This observation motivated the final modification.

### 2.3.4 In-memory storage via tmpfs (RamDisk)

To reduce the I/O bottleneck, intermediate data and temporary files are stored in the temporary file system (tmpfs), a memory-backed file system in Linux. This approach reduces disk access during intermediate I/O operations and reduces competition for disk bandwidth among containers running simultaneously. In our tests, this configuration, used on computing instances with large amounts of memory, achieved shorter runtimes than a configuration using high-speed non-volatile memory express (NVMe) solid-state storage.

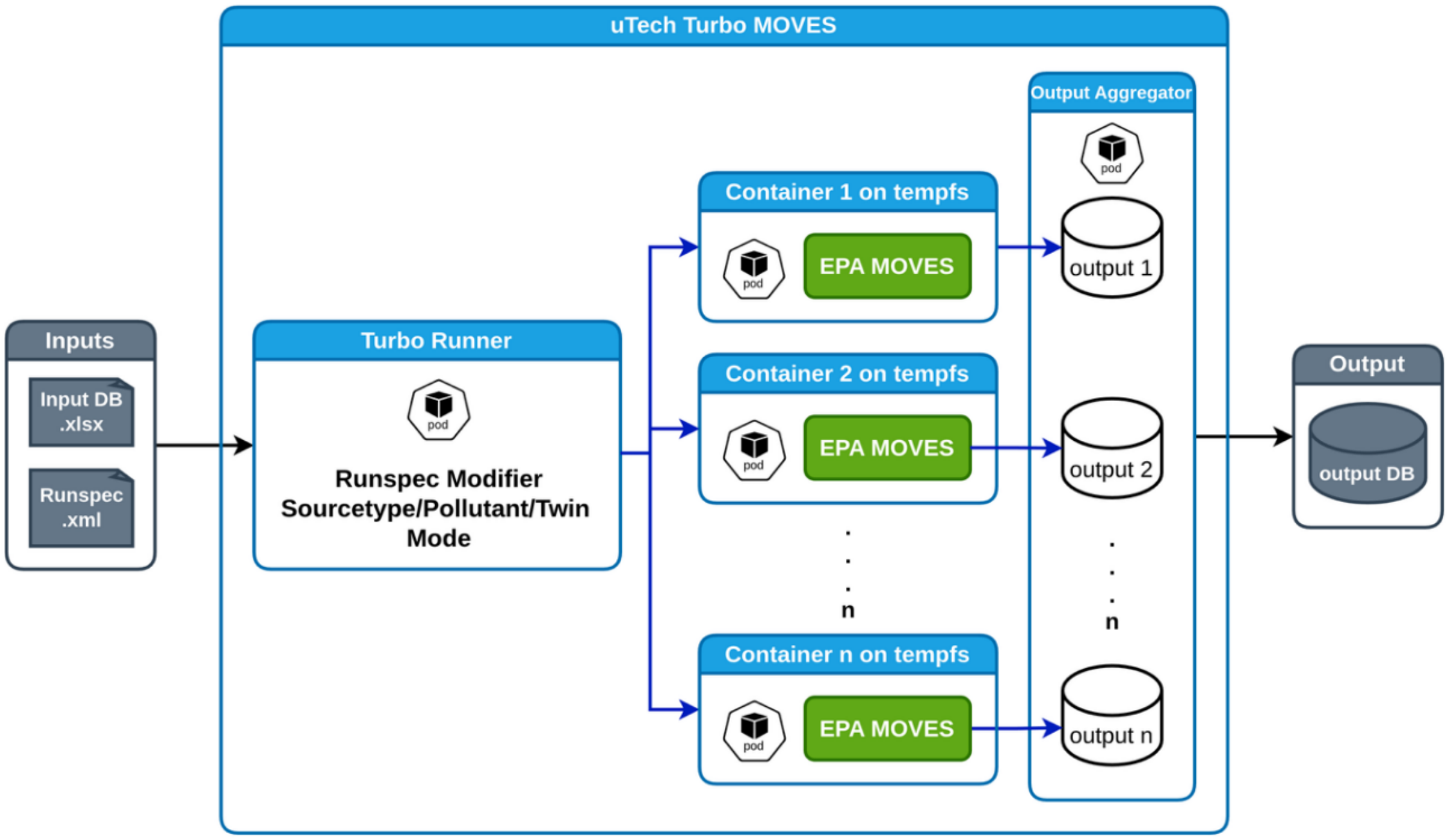

**Figure 3.** uTECH Turbo-MOVES Computation Architecture

### 2.3.5 Integrated workflow and cloud deployment

Fig. 3 summarizes the end-to-end workflow. The Turbo Runner ingests a domain input database and a RunSpec, partitions the RunSpec according to the selected splitting dimension, and dispatches each sub-run to a dedicated tmpfs-backed container, with the containers executing concurrently on a single machine. Each container produces an isolated output, and the Output Grouper reassembles these outputs into a single result equivalent to that of an unsplit run, summing partial mass totals in Inventory mode and taking the union of rate rows in Emission Rates mode.

Fig. 4 illustrates the corresponding multi-run deployment on Amazon Web Services. Per-county input files are staged in S3, independent EC2 instances execute uTECH Turbo-MOVES for their assigned counties concurrently, and results are consolidated in a central RDS database. Because every component of the architecture alters deployment rather than computation, output equivalence with default MOVES is expected by construction.

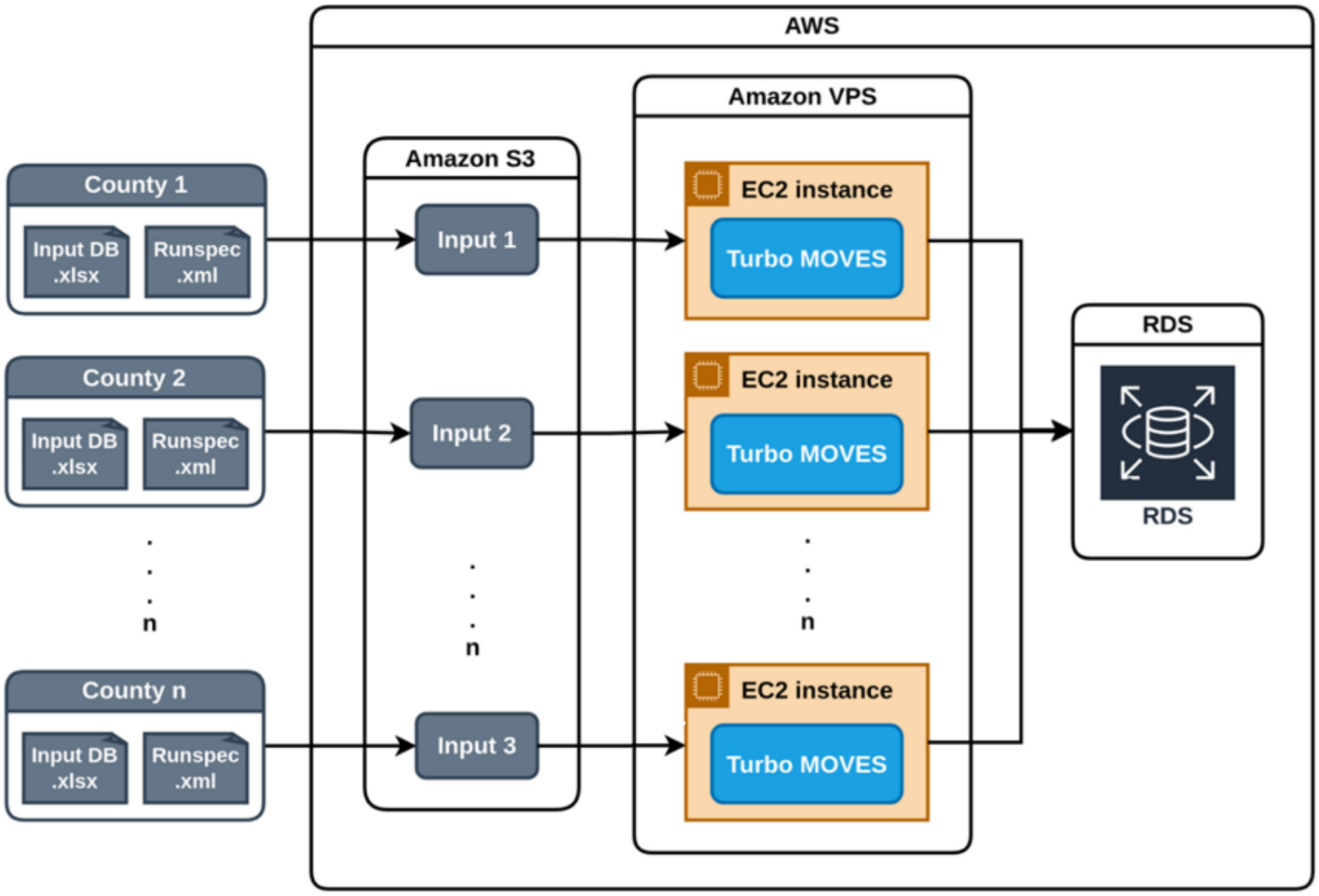

**Figure 4.** Multi-instance cloud deployment of uTECH Turbo-MOVES.

Beyond a single machine, campaigns comprising many independent runs are sized across multiple concurrently dispatched instances rather than executed on one. A capacity-planning estimator takes the RunSpecs of a batch together with a target machine's memory, vCPU, and disk throughput budget and, using each case's measured per-run resource footprint, computes the number of machines and dispatch waves required to complete the batch. The instance counts used for the Project-scale campaigns in Section 3.2 follow this estimator's sizing rather than any particular ad hoc dispatch.

## 2.4. Experimental design

The evaluation is organized around the central claim of this paper, that uTECH Turbo-MOVES accelerates any EPA MOVES analysis while returning outputs identical to those of the default model. Two principal configuration choices define a MOVES analysis: the scale of analysis and the calculation type. All three scales draw on the same underlying emission rates and adjustment factors, and either calculation type may be combined with any scale. These two axes span a 3×2 configuration space that covers the model's principal applications. County scale is required for State Implementation Plan and regional transportation conformity analyses, Project scale is required for project-level hot-spot conformity determinations, and Default scale supports national and state-level screening and greenhouse gas inventory development.

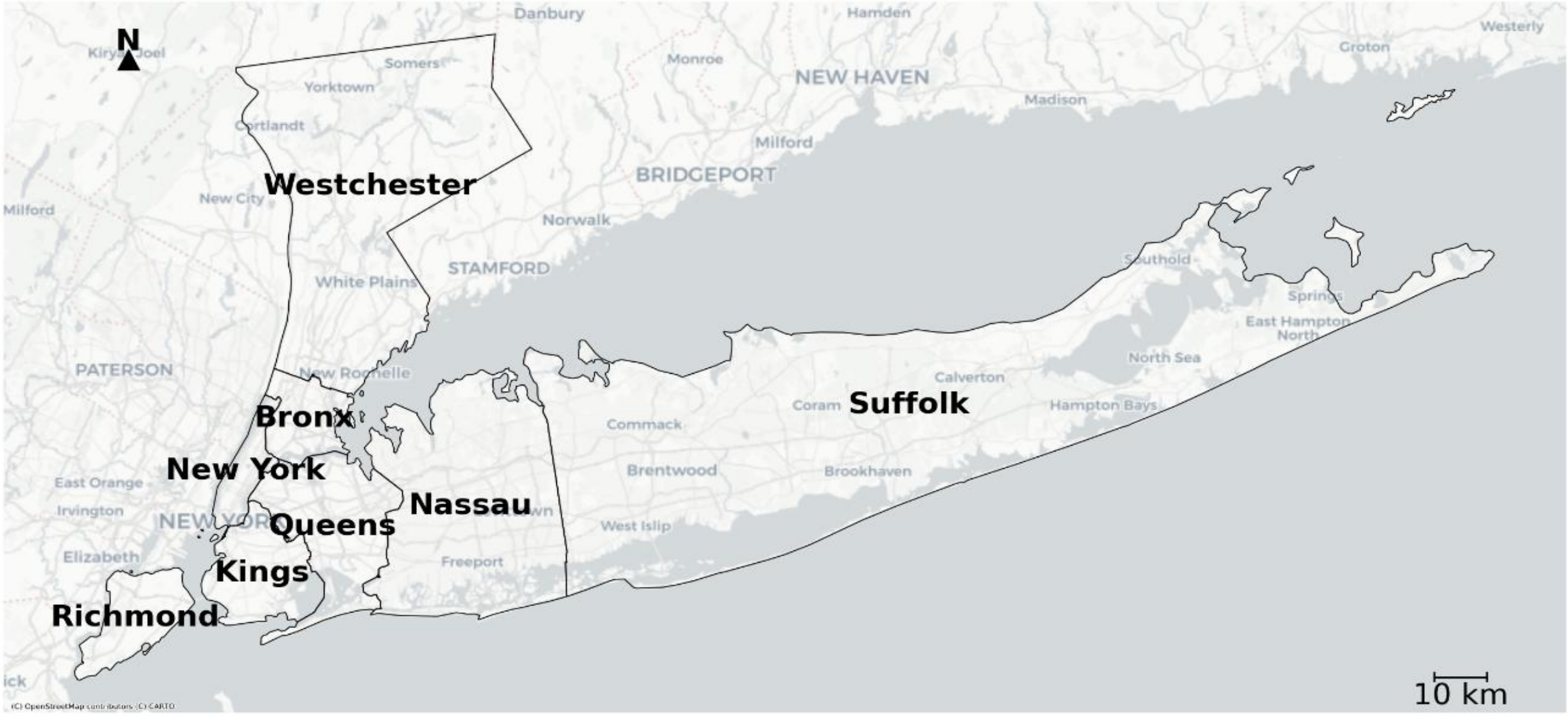

**Figure 5.** The eight New York Metropolitan Transportation Council counties

Rather than benchmarking a single configuration, we evaluate uTECH Turbo-MOVESin all six cells of this configuration space, instantiating each cell as a canonical regulatory use case. Geography is held constant across the County- and Default-scale cases at the eight-county New York Metropolitan Transportation Council region (Fig. 5), and the Project-scale cases share a single highway interchange, so that within each scale the only factor varying between cases is the MOVES configuration itself. Because the architecture modifies only how a run is partitioned and deployed, demonstrated coverage of the full configuration space supports the generality of the approach.

### 2.4.1. Benchmark cases

Table 2 summarizes the six benchmark cases and the mechanism for each exercise. The County-scale Inventory and Emission Rates cases used input data from the New York MOVES5 Input Dataset (Department of Environmental Conservation, 2026).

**Table 2** Benchmark case matrix spanning the EPA MOVES configuration space.

| Case | Scale × Type | Geography | Year | Temporal span | Pollutants |
|---|---|---|---|---|---|
| Case A | County × Inventory | NYMTC counties | 2023 | Winter and summer; weekday | NOx; VOC |
| Case B | County × Rates | NYMTC counties | 2019 | Winter; weekday and weekend | CO2eq; Total energy |
| Case C | Default × Inventory | NYMTC counties | 2025 | Annual; weekday and weekend | CO2eq; Total energy |
| Case D | Project × Inventory | 15 directional links | 2027 | Winter; 24 hourly RunSpecs | PM2.5; NOx |
| Case E | Project × Rates | 15 directional links | 2027 | Winter; 24 hourly RunSpecs | PM2.5; NOx |
| Case F | Default × Rates | NYMTC counties | 2025 | Annual; weekday | CO2eq; Total energy |

*Note. Cases D and E each comprise 48 RunSpecs (24 h × 2 scenarios).*

Case A comprises County-scale Inventory runs for the eight counties, mirroring the structure of a regional conformity analysis, with one winter and one summer weekday, hourly activity and hourly output, NOx and VOC, all on-road source types with their applicable fuel types, and all five MOVES road types. The summer day requires month-specific meteorology and fuel inputs supplied through the County Data Manager. The baseline executes the eight county RunSpecs sequentially. uTECH Turbo-MOVES combines inter-run parallelism across the eight county RunSpecs with intra-run splitting of each RunSpec by source type, which is used in every uTECH Turbo-MOVES run.

Case B is the canonical benchmarking configuration, producing hourly emission rates for 2 and 5 January 2019 for the energy and greenhouse gas pollutant set. Runs span all thirteen onroad source use types with their applicable fuel types, all emission processes, and all five MOVES road types, with outputs disaggregated by model year, fuel type, process, road type, source use type, source classification code, and regulatory class, alongside activity outputs. Emission Rates mode is the pathway used by SMOKE-MOVES-based air quality modeling platforms and representative-county workflows, and it produces the model's complete rate-table set, making it the longest-running and most output-intensive calculation type.

Case C is a single Default-scale Inventory RunSpec covering the eight-county region for calendar year 2025, all twelve months and both day types, for carbon dioxide equivalent and total energy using the MOVES default database. It is the largest single RunSpec in the suite and the principal test of strong scaling under intra-run splitting.

Case D represents a single highway interchange as fifteen directional links specified through the Project Data Manager. Consistent with hot-spot guidance, one RunSpec is required per modeled hour (US EPA, 2016), so the campaign comprises 24 hourly RunSpecs for each of two scenarios, that is, 48 Project-scale Inventory runs for particulate matter and nitrogen oxides in a winter month. A full four-season analysis multiplies the campaign by four and scales identically.

Cases E and F complete the configuration space. Case E mirrors Case D in Emission Rates mode, comprising 24 hourly RunSpecs for each of the build and no-build scenarios, that is, 48 Project-scale runs producing the link-level rate tables consumed by dispersion models such as AERMOD.

Case F produces Default-scale distance-based screening rates for the eight-county region for weekdays in all twelve months of calendar year 2025.

#### 2.4.2. Acceleration regimes and performance metrics

The architecture accelerates MOVES through two distinct mechanisms, and the cases are chosen to exercise both. Under intra-run splitting, a single large RunSpec is partitioned by source type, pollutant, or both into independent sub-runs that execute concurrently in isolated containers, and the Output Grouper reassembles a single result. Under inter-run parallelism, many small RunSpecs, which standard MOVES must execute sequentially because the Main program binds port 13131, run simultaneously as independent containers.

The two regimes call for different performance metrics. We report per-run speedup for intra-run splitting and campaign completion time, calculated as the wall-clock time to complete a defined batch, for inter-run parallelism. A single small Project-scale run already completes in minutes under default MOVES and gains little from acceleration; the practical burden at Project scale arises instead from guidance requiring a separate RunSpec for each modeled hour, so that a quantitative hot-spot analysis comprises dozens of runs. Cases D and E are therefore specified, executed, and reported as batch campaigns, in Inventory and Emission Rates mode respectively.

Execution times for the benchmark cases were calculated from the start and finish timestamps recorded in the execution logs for each completed run or campaign. Speedup was calculated as the ratio of the corresponding baseline execution time to the uTECH Turbo-MOVES execution time. The multi-county scaling estimates in Section 3.3 were derived separately from the mean per-county runtimes.

Output equivalence was assessed by comparing the movesoutput tables produced by the three configurations. For each county run, output rows were grouped by pollutant and emission process, and the number of rows and the summed emission quantity in each group were compared across the Windows, Linux, and uTECH Turbo-MOVES runs; for uTECH Turbo-MOVES, the group totals were computed from the combined outputs of the thirteen sub-runs. Outputs were considered equivalent when row counts matched exactly and emission totals differed only at the level of floating-point rounding. The comparison script, the grouped summaries for every county, and the complete raw output databases for New York County are provided in the repository referenced in the Data Availability section, so that the comparison can be reproduced without the full set of output databases.

### 2.4.3. Benchmark configurations and execution environment

We compare three system configurations. The first is the standard EPA MOVES distribution on Windows, configured with one Main and three Worker processes following Table 1; this represents the typical practitioner environment and serves as the primary baseline. The second is the same setup in a Linux environment, representing the improved baseline. The third is uTECH Turbo-MOVES, comprising containerized execution with in-memory storage, RunSpec splitting, and parallel orchestration.

Baseline configurations execute on storage-optimized Amazon EC2 i7i.4x large instances with 16 vCPU, 128 GiB of memory, and local NVMe solid-state storage, a deliberate choice that grants the disk-bound baseline its most favorable storage subsystem. uTECH Turbo-MOVES executes on m7a.16xlarge instances with 64 vCPU and 256 GiB of memory, whose capacity is dictated by the memory demands of tmpfs-resident execution; preliminary profiling on a 52-core workstation showed peak memory use of approximately 92 GB for a single county-level run and informed this sizing. To exclude instance effects from the comparison, the Linux baseline is additionally executed on the m7a.16xlarge instance for the canonical case.

All experiments use EPA MOVES 5.0.1 with MariaDB 11.4 and Docker 20.10.24. All RunSpecs, input databases, and run logs are archived by run identifier and provided with the accompanying code in the GitHub repository referenced in the Data Availability section.

## 3. Results

All comparisons in this section are between the deployed systems described in Section 2.4.3, which differ in instance configuration and, for campaigns, in the number of concurrently used instances, as well as in software architecture. The reported speedups are therefore neither equal-resource nor resource-normalized comparisons, and they do not isolate the contribution of individual architectural changes under equal hardware allocation.

### 3.1. End-to-end performance on the canonical case

To establish the end-to-end speedup, we executed the canonical Case B configuration for all eight counties under the three system configurations defined in Section 2.4.3. Table 3 reports the per-county execution times, together with the corresponding times on a local Windows workstation as a reference for the typical practitioner environment.

**Table 3.** Execution times for the eight NYMTC county runs in the Case B configuration (County-scale Emission Rates) under the NY Dataset.[2]

| County | Workstation, Windows (min) | Cloud, Windows (min) | Cloud, Linux (min) | uTECH Turbo-MOVES (min) | Speedup |
|---|---|---|---|---|---|
| Bronx | 179.4 | 185.4 | 127.5 | 16.2 | 11.4× |
| Kings | 186.4 | 177.6 | 131.2 | 13.6 | 13.1× |
| Nassau | 181.0 | 168.5 | 127.0 | 13.4 | 12.6× |
| New York | 184.1 | 192.5 | 135.0 | 17.0 | 11.3× |
| Queens | 191.9 | 184.6 | 125.6 | 13.6 | 13.6× |
| Richmond | 201.5 | 188.5 | 129.1 | 13.1 | 14.4× |
| Suffolk | 194.5 | 174.0 | 122.6 | 13.5 | 12.9× |
| Westchester | 187.7 | 170.0 | 125.4 | 13.3 | 12.8× |
| Mean | 188.3 | 180.1 | 127.9 | 14.2 | 12.7× |
| Campaign wall-clock | 1,506.5 | 1,441.1 | 1,023.4 | 17.0 | 84.8× |

[2] Runtimes are log-derived and reported in minutes. Speedup = cloud Windows runtime / uTECH-Turbo MOVES runtime; the mean-row value uses the ratio of mean runtimes. In the campaign wall-clock row, baseline campaign times are the sums of the eight sequentially executed county runs, and the uTECH-Turbo MOVES campaign time is that of the slowest of the eight concurrently executed counties.

The calculated mean runtime was 188.3 min per county for the local Windows workstation, 180.1 min for the cloud Windows configuration, and 127.9 min for the cloud Linux configuration. The cloud Linux runtime was lower than the cloud Windows runtime by a factor of 1.4. uTECH Turbo-MOVES completed the same county workloads in a mean of 14.2 min, corresponding to speedups of 12.7× relative to cloud Windows and 9.0× relative to cloud Linux.

Per-county runtimes under uTECH Turbo-MOVES ranged from 13.1 min in Richmond to 17.0 min in New York. This limited variation is consistent with similar computational demands across the configured county workloads, but does not by itself establish which workload characteristics determine runtime. For Case B, Table 4 compares sequential baseline campaigns of 1,441 min on Windows and 1,023 min on Linux with a concurrent uTECH Turbo-MOVES campaign of 17.0 min. The resulting Windows-relative campaign speedup of 84.8× combines lower per-county runtimes with concurrent execution of the eight counties. If the baseline county runs were instead dispatched simultaneously on separate instances, their longest reported runtimes would be 192.5 min on Windows and 135.0 min on Linux, before allowing for dispatch and consolidation overhead. These conditional estimates are distinct from the sequential campaign comparison in Table 4.

### 3.2. Performance across the configuration space

The six benchmark cases cover the combinations of MOVES scale and calculation type specified in Section 2.4.1. Table 4 reports calculated campaign completion times for Cases A, B, D, and E and individual RunSpec execution times for Cases C and F. The corresponding ratios are therefore campaign speedups and per-run speedups, respectively, and should be read separately.

Under inter-run parallelism, Case A mirrors a regional conformity analysis. The Windows baseline executed the eight conformity RunSpecs sequentially in 53.25 min, a mean of 6.66 min per run, and the Linux baseline in 36.57 min, whereas uTECH Turbo-MOVES executed them as a single parallel batch, so that the campaign time equals that of the slowest county, 7.02 min. The resulting campaign speedup was 7.6× relative to Windows and 5.2× relative to Linux. Because the uTECH Turbo-MOVES per-county time exceeded the baseline mean per-run time, this benefit arose from concurrent execution of independent RunSpecs rather than faster individual runs.

Case C, the annual Default-scale Inventory RunSpec, was partitioned by source type into thirteen independent sub-runs. Its calculated runtime decreased from 91.45 min under the Windows baseline and 93.55 min under the Linux baseline to 16.5 min under uTECH Turbo-MOVES, corresponding to per-run speedups of 5.5× and 5.7×, respectively. Dividing the Windows-relative speedup by the thirteen sub-runs gives 42%, but this ratio is not a controlled measure of parallel efficiency. The resource profiles in Fig. 6 are consistent with reduced disk activity and greater CPU utilization under uTECH Turbo-MOVES. They do not isolate the effects of partitioning, in-memory storage, and hardware configuration. The peak memory increase, measured in a dedicated profiling run of the same RunSpec, was 34.7 GB.

**Table 4.** Performance across the six cells of the MOVES configuration space.

| Case | Scale × Type | Runs | Metric[3] | MOVES, Windows (mins) | MOVES, Linux (mins) | uTECH Turbo-MOVES (mins) | Speedup |
|---|---|---|---|---|---|---|---|
| A | County × Inventory | 8 | Campaign wall-clock | 53.25 | 36.57 | 7.02 | 7.6× |
| B | County × Rates | 8 | Campaign wall-clock | 1,441 | 1,023 | 17.0 | 84.8× |
| C | Default × Inventory | 1 | Single-run time | 91.45 | 93.55 | 16.5 | 5.5× |
| D | Project × Inventory | 48 | Campaign wall-clock | 101.12 | 86.78 | 4.90 | 20.6× |
| E | Project × Rates | 48 | Campaign wall-clock | 99.97 | 79.2 | 5.10 | 19.6× |
| F | Default × Rates | 1 | Single-run time | 3,570[a] | 2,538[a] | 233.7 | 15.3×[a] |

[3] All execution times are reported in minutes. Runs are counts, and speedups are dimensionless ratios relative to MOVES on Windows. Baseline campaign times are the sums of sequentially executed runs; uTECH-Turbo MOVES campaign times are the completion times of concurrently dispatched runs. Values marked [a] are estimated.

Under inter-run parallelism, Case D comprised 48 Project-scale Inventory runs. The calculated sequential campaign times were 101.12 min for the Windows baseline and 86.78 min for the Linux baseline; the Windows mean per-run time was 2.11 min. uTECH Turbo-MOVES completed the concurrent campaign in 4.90 min, corresponding to campaign speedups of 20.6× and 17.7×, respectively. The deployment used six instances providing 48 concurrent execution slots; measured peak memory use was 11.9 to 14.0 GB per run, and CPU use averaged 3.0 vCPU per run. Individual uTECH Turbo-MOVES runs took 4.6 to 4.9 min, longer than the Windows mean of 2.11 min. The campaign benefit therefore arose from concurrent dispatch rather than faster individual execution.

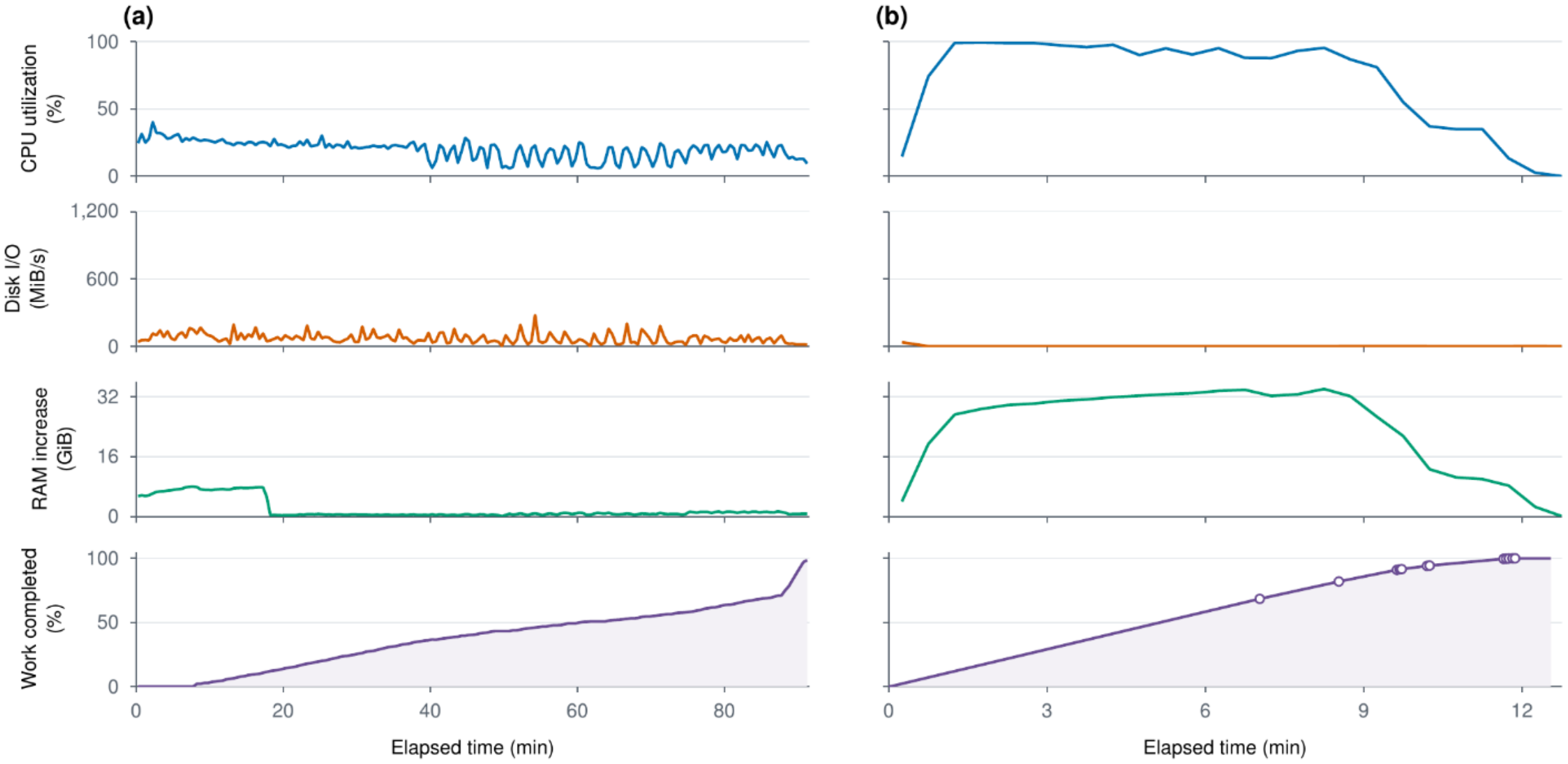

**Fig. 6.** CPU utilization, disk throughput, memory increase, and work completed for Case C under (a) EPA MOVES on Windows with one Main and three Workers and (b) uTECH Turbo-MOVES with thirteen parallel sub-runs, measured in a dedicated profiling run of the same RunSpec (11.85 min). Lines show 30 s means; in (b), work completed is cumulative sub-run time as a share of the total, so its slope is proportional .to the number of sub-runs still running, and markers show sub-run completions

Case E comprised the corresponding 48-run Project-scale Emission Rates campaign. The calculated sequential campaign times were 99.97 min for the Windows baseline and 79.2 min for the Linux baseline; the Windows mean per-run time was 2.08 min. uTECH Turbo-MOVES dispatched the 48 runs concurrently across six instances, as in Case D, and completed the campaign in 5.10 min, corresponding to campaign speedups of 19.6× relative to Windows and 15.5× relative to Linux. As in Case D, individual runs took 4.7 to 5.1 min, longer than the Windows mean of 2.08 min, so the campaign benefit again arose from concurrent dispatch.

Case F produced Default-scale distance-based screening rates for the eight-county region for weekdays in all twelve months of 2025. Its calculated runtime was 233.7 min under uTECH Turbo-MOVES, compared with an estimated 59.5 h (3,570 min) for the Windows baseline, corresponding to an estimated per-run speedup of 15.3×. The reported peak memory use was 176.4 GB, the highest value in the suite.

**Table 5. Output equivalence for Case A across EPA MOVES on Windows, EPA MOVES on Linux, and uTECH Turbo-MOVES.**

| County | Groups[a] | Rows compared | Row counts match | Max. relative difference[b] |
|---|---|---|---|---|
| Bronx | 7 | 534,816 | Yes | 1.42 |
| Kings | 7 | 534,816 | Yes | 2.13 |
| Nassau | 7 | 568,512 | Yes | 1.19 |
| New York | 7 | 534,816 | Yes | 1.49 |
| Queens | 7 | 534,816 | Yes | 1.89 |
| Richmond | 7 | 534,816 | Yes | 1.79 |
| Suffolk | 7 | 568,512 | Yes | 1.04 |
| Westchester | 7 | 554,544 | Yes | 2.04 |
| All counties | 56 | 4,365,648 | Yes | 2.13 |

[a] *Pollutant and emission-process combinations in the movesoutput table.*

[b] *Largest difference in a group's summed emission quantity among the three configurations, divided by the largest value, in units of* $10^{-9}$.

### 3.3. Scaling projections for multi-county campaigns

Table 6 presents scaling projections rather than runtimes calculated from completed multi-county campaigns. Total machine time is estimated by multiplying the county count by the mean of the per-county runtimes for the corresponding Case B configuration in Table 3. This linear extrapolation assumes that the benchmark mean represents the additional counties and that each county is executed on one instance of the corresponding configuration. It does not account for differences in county workloads, orchestration overhead, or resource contention. Machine-hours describe instance occupation time, not hardware-normalized computational work.

**Table 6.** Projected aggregate machine time for multi-county analyses under the default EPA MOVES distribution (Windows and Linux) and uTECH Turbo-MOVES.

| Analysis type | County count | Sequential workstation time (hours) | Cloud Windows (machine-h) | Cloud Linux (machine-h) | uTECH Turbo-MOVES (machine-h) |
|---|---|---|---|---|---|
| Individual county | 1 | 3.12 | 3.0 | 2.1 | 0.24 |
| New York State | 62 | 194.4 | 186.1 | 132.2 | 14.7 |
| Texas | 254 | 796.4 | 762.4 | 541.4 | 60.1 |
| Representative counties | 300 | 940.8 | 900.5 | 639.5 | 71.0 |
| Midwest (12 states) | 1,055 | 3,312.0 | 3,166.8 | 2,248.9 | 249.7 |
| Whole US* | 3,000+ | 9,415.2 | 9,005 | 6,395 | 710 |

** The United States has approximately 3,143 counties and county equivalents; the projection uses 3,000.*

The national-scale row in Table 6 projects approximately 9,000 machine-hours under the cloud Windows baseline, 6,400 machine-hours under the Linux baseline, and 710 machine-hours under uTECH Turbo-MOVES. Sequential execution of the national campaign with the default distribution on a single workstation is estimated to require 9,415.2 hours under this linear extrapolation. Dividing the projected machine-hours by 100 concurrent instances gives an idealized completion time of approximately 7 h for uTECH Turbo-MOVES, compared with roughly 91.2 hours for the default Windows distribution at the same level of concurrency. These estimates assume full utilization, balanced workloads, and negligible additional overhead. They are not demonstrated national campaign runtimes.

## 4. Discussion

The findings demonstrate that changes to execution architecture can reduce the computational burden of MOVES without replacing its Calculator modules. By combining workload partitioning, container isolation, in-memory storage, and concurrent dispatch, uTECH Turbo-MOVES addresses constraints associated with both individual analyses and collections of related runs. The contribution is therefore an alternative execution strategy for existing MOVES workflows, rather than a new emissions model. Its practical significance lies in making repeated analyses more feasible as transport activity, fleet composition, and scenario assumptions change.

Intra-run splitting is applied to every uTECH Turbo-MOVES run. For campaigns comprising multiple independent RunSpecs, inter-run parallelism provides an additional level of concurrency. These mechanisms have distinct implications for interpreting performance. A shorter campaign completion time may reflect concurrent execution as well as changes in individual runtimes, and does not imply an equivalent reduction in the computational work required for each run. The Project-scale benchmarks illustrate this distinction, with shorter campaign times despite longer individual runtimes. Performance should therefore be interpreted in relation to workload structure and the level at which execution time is measured.

### 4.1. Implications for practice

Regional transportation conformity analysis illustrates the practical importance of reducing the time required for repeated emissions calculations. EPA guidance identifies separate county inventories as an approach to multi-county analysis, notes the burden of requiring an individual run for each county, and emphasizes the use of current planning inputs (US EPA, 2024a). The findings for Case A suggest that faster campaign execution could shorten the interval between revising these inputs and reviewing regional emissions estimates. The benefit is therefore a more manageable process for updating analyses, rather than faster execution of an isolated model run.

Emission-rate generation creates a related computational burden within emissions-inventory and air-quality modeling workflows. EPA identifies longer runtimes and larger output files as considerations for Emission Rates mode and describes the use of MOVES rate tables in SMOKE-based emissions processing (US EPA, 2024a). Case B illustrates the potential to shorten rate-table generation when revised inputs require new calculations. Because this benchmark uses a greenhouse-gas and energy configuration, its results should not be interpreted as performance measurements for a complete air-quality modeling application. The practical opportunity is to reduce the time required for an upstream computational task.

Regional greenhouse-gas assessment provides a further opportunity for repeated scenario evaluation. EPA guidance identifies Default-scale modeling as useful for preliminary screening and selected comparative analyses, including comparisons across vehicle types, operating conditions, and future scenarios (US EPA, 2024b). The shorter individual-run times observed in Cases C and F suggest that sensitivity analysis and scenario refinement could become more feasible within a fixed analytical schedule. This would expand the range of assumptions that practitioners can examine, rather than improve emissions-estimation accuracy through faster computation alone.

At the project level, the relevant workload often comprises multiple calculations for alternative designs and operating conditions. EPA hot-spot guidance describes build and no-build comparisons and specifies that each Project-scale MOVES run represents a single modeled hour (US EPA, 2021). Cases D and E show why the time required to complete the full set of calculations is a useful performance measure. Inter-run parallelism, applied alongside intra-run splitting, could enable earlier preparation of emissions inputs for dispersion modeling and facilitate repeated evaluation of project alternatives. These opportunities depend on available computing resources and do not eliminate the time required for data preparation, quality assurance, or interpretation.

## 5. Conclusions and future directions

This study presented uTECH Turbo-MOVES, an execution architecture that combines RunSpec partitioning, container isolation, in-memory storage, and concurrent dispatch while retaining the MOVES Calculator modules. Across six benchmark configurations, the evaluated deployments reduced execution times for large individual RunSpecs and shortened the completion of multi-run campaigns. The two Default-scale benchmarks achieved per-run speedups of 5.5× (measured) and 15.3× (estimated), while the Project-scale campaigns demonstrated the benefit of inter-run parallelism despite longer individual runtimes.

These findings establish execution architecture as a practical means of reducing the computational burden of repeated MOVES analyses. The resulting opportunity is to support more extensive scenario evaluation within available analytical schedules. The reported gains remain specific to the tested workloads and deployment configurations, rather than a universal acceleration factor or a resource-normalized measure of efficiency.

Future evaluation should prioritize repeated benchmarks under matched hardware and resource budgets, followed by controlled tests that vary partitioning, in-memory storage, and concurrency separately. Larger multi-county campaigns would test the scaling projections and quantify scheduling and consolidation overhead. Workflow extensions could then connect uTECH Turbo-MOVES to upstream transportation activity models and evaluate whether shorter model execution reduces total analytical turnaround. Extending the approach to MOVES-Nonroad is another candidate application, subject to separate implementation and testing.

## Data Availability

The experimental input files, run configurations, run logs, and the scripts that recompute the reported execution times from those logs and the Case A output-equivalence comparison are available in the uTECH Turbo-MOVES GitHub repository (https://github.com/caferavci/uTECH Turbo-MOVES). The underlying New York State MOVES input dataset is publicly available through the New York State Open Data portal, with the source linked in the repository.

## Declaration of Competing Interest

The authors declare that there are no financial and personal relationships with other people or organizations that can inappropriately influence our work.

## CRediT Authorship Contribution Statement

**Cafer Avci:** Writing – original draft, Writing – review & editing, Validation, Software, Methodology, Data curation, Project administration, Conceptualization. **Alireza Yazdiani:** Writing – original draft, Writing – review & editing, Data curation, Software, Methodology, Validation. **Roopak Srinivasan:** Writing – original draft, Software, Methodology, Data curation, Validation. **Megh Khaire:** Software, Methodology, Data curation. **Mohammad Tayarani:** Writing – review & editing. **H. Oliver Gao:** Writing – review & editing, Supervision, Resources, Conceptualization.

## References

Allcott, H., Kane, R., Maydanchik, M.S., Shapiro, J.S., Tintelnot, F., 2024. The effects of "buy american": Electric vehicles and the inflation reduction act. National Bureau of Economic Research.

Barber, R., Lohman, G., Pandis, I., Raman, V., Sidle, R., Attaluri, G., Chainani, N., Lightstone, S., Sharpe, D., 2014. Memory-efficient hash joins. Proc. VLDB Endow. 8, 353–364. https://doi.org/10.14778/2735496.2735499

Bui, A., Yang, Z., 2022. U.S. light-duty vehicle greenhouse gas standards for model years 2023–2026 and corporate average fuel economy standards for model years 2024–2026.

Cook, C., Kreidieh, A., Vasserman, S., Allcott, H., Arora, N., van Sambeek, F., Tomkins, A., Turkel, E., 2025. The Short-Run Effects of Congestion Pricing in New York City. National Bureau of Economic Research.

Department of Environmental Conservation, 2026. MOVES5 2024 Input Data Tables | State of New York [WWW Document]. MOVES5 2024 Input Data Tables. URL https://data.ny.gov/Energy-Environment/MOVES5-2024-Input-Data-Tables/ybv6-5sxa/about_data (accessed 10.10.26).

Fraser, T., Guo, Y., Gao, H.O., 2024. Making MOVES move: Fast emissions estimates for repeated transportation policy scenario analyses. Environ. Model. Softw. 178, 106084. https://doi.org/10.1016/j.envsoft.2024.106084

Fujita, E.M., Campbell ,David E., Zielinska ,Barbara, Chow ,Judith C., Lindhjem ,Christian E., DenBleyker ,Allison, Bishop ,Gary A., Schuchmann ,Brent G., Stedman ,Donald H., and Lawson, D.R., 2012. Comparison of the MOVES2010a, MOBILE6.2, and EMFAC2007 mobile source emission models with on-road traffic tunnel and remote sensing measurements. J. Air Waste Manag. Assoc. 62, 1134–1149. https://doi.org/10.1080/10962247.2012.699016

Gu, Z., Liu, Z., Cheng, Q., Saberi, M., 2018. Congestion pricing practices and public acceptance: A review of evidence. Case Stud. Transp. Policy 6, 94–101. https://doi.org/10.1016/j.cstp.2018.01.004

Hall-Quinlan, D.L., He, H., Ren, X., Canty, T.P., Salawitch, R.J., Stratton, P., Dickerson, R.R., 2023. Inferred vehicular emissions at a near-road site: Impacts of COVID-19 restrictions, traffic patterns, and ambient air temperature. Atmos. Environ. 299, 119649. https://doi.org/10.1016/j.atmosenv.2023.119649

Khan, R.N., Saporito ,Antonio F., Zenon ,Jania, Goodman ,Lael, and Zelikoff, J.T., 2024. Traffic-related air pollution in marginalized neighborhoods: a community perspective. Inhal. Toxicol. 36, 343–354. https://doi.org/10.1080/08958378.2024.2331259

Liu, H., Guensler ,Randall, Lu ,Hongyu, Xu ,Yanzhi, Xu ,Xiaodan, and Rodgers, M.O., 2019. MOVES-Matrix for high-performance on-road energy and running emission rate modeling applications. J. Air Waste Manag. Assoc. 69, 1415–1428. https://doi.org/10.1080/10962247.2019.1640806

Liu, H., Xu, X., Rodgers, M.O., Xu, Y. "Ann", Guensler, R.L., 2017. MOVES-Matrix and distributed computing for microscale line source dispersion analysis. J. Air Waste Manag. Assoc. 67, 763–775. https://doi.org/10.1080/10962247.2017.1287788

Nadafianshahamabadi, R., 2019. Is Transportation Planning Effective? A Critical Review of Long-range Regional Transportation Planning in the United States (Ph.D.). The University of New Mexico, United States -- New Mexico.

NHTSA, 2025. Corporate Average Fuel Economy [WWW Document]. URL https://www.nhtsa.gov/laws-regulations/corporate-average-fuel-economy (accessed 6.27.25).

Ramirez-Sanchez, E., Tang, C., Xu, Y., Renganathan, N., Jayawardana, V., He, Z., Wu, C., 2025. NeuralMOVES: A lightweight and microscopic vehicle emission estimation model based on reverse engineering and surrogate learning. https://doi.org/10.48550/arXiv.2502.04417

StandfordReport, 2024. Electric vehicle subsidies help the climate and automakers, but at questionable cost to taxpayers [WWW Document]. StandfordReport. URL https://news.stanford.edu/stories/2024/10/electric-vehicle-subsidies-help-the-climate-u-s-automakers

Tayarani, M., Poorfakhraei, A., Nadafianshahamabadi, R., Rowangould, G., 2018. Can regional transportation and land-use planning achieve deep reductions in GHG emissions from vehicles? Transp. Res. Part Transp. Environ. 63, 222–235. https://doi.org/10.1016/j.trd.2018.05.010

Tayarani, M., Poorfakhraei, A., Nadafianshahamabadi, R., Rowangould, G.M., 2016. Evaluating unintended outcomes of regional smart-growth strategies: Environmental justice and public health concerns. Transp. Res. Part Transp. Environ. 49, 280–290. https://doi.org/10.1016/j.trd.2016.10.011

US EPA, O., 2016. Project-Level Conformity and Hot-Spot Analyses [WWW Document]. URL https://www.epa.gov/state-and-local-transportation/project-level-conformity-and-hot-spot-analyses (accessed 10.10.26).

US EPA, 2021. PM Hot-Spot Guidance: Transportation Conformity Guidance for Quantitative Hot-spot Analyses in PM2.5 and PM10 Nonattainment and Maintenance Areas. EPA-420-B-21-037. Office of Transportation and Air Quality, U.S. Environmental Protection Agency, October 2021.

US EPA, 2024a. MOVES5 Technical Guidance: Using MOVES to Prepare Emission Inventories for State Implementation Plans and Transportation Conformity. EPA-420-B-24-043. Office of Transportation and Air Quality, U.S. Environmental Protection Agency, November 2024.

US EPA, 2024b. MOVES Greenhouse Gas Guidance: Using MOVES for Estimating State and Local Inventories of Onroad Greenhouse Gas Emissions and Energy Consumption. EPA-420-B-24-023. Office of Transportation and Air Quality, U.S. Environmental Protection Agency, June 2024.

US EPA, 2025a. Transportation Sector Emissions [WWW Document]. URL https://www.epa.gov/ghgemissions/transportation-sector-emissions (accessed 6.27.25).

US EPA, 2025b. Latest Version of MOtor Vehicle Emission Simulator (MOVES) [WWW Document]. URL https://nepis.epa.gov/Exe/ZyPDF.cgi?Dockey=P101DQ9V.pdf (accessed 10.10.26).

Wang, Z., Wu, G., Scora, G., 2021. MOVESTAR: An Open-Source Vehicle Fuel and Emission Model based on USEPA MOVES. https://doi.org/10.48550/arXiv.2008.04986

Winkless, L., 2025. Congestion Pricing Can Improve Air Quality And Human Health In Cities [WWW Document]. Forbes. URL https://www.forbes.com/sites/lauriewinkless/2025/03/11/congestion-pricing-can-improve-air-quality-and-human-health-in-cities/ (accessed 6.27.25).

Wu, D., Zheng, A., Yu, W., Cao, H., Ling, Q., Liu, J., Zhou, D., 2025. Digital Twin Technology in Transportation Infrastructure: A Comprehensive Survey of Current Applications, Challenges, and Future Directions. Appl. Sci. 15, 1911. https://doi.org/10.3390/app15041911

Xu, X., Liu, H., Anderson, J.M., Xu, Y. (Ann), Hunter, M.P., Rodgers, M.O., Guensler, R.L., 2016. Estimating Project-Level Vehicle Emissions with Vissim and MOVES-Matrix. Transp. Res. Rec. J. Transp. Res. Board 2570, 107–117. https://doi.org/10.3141/2570-12

Zhao, Y., Sadek, A.W., 2013. Computationally-efficient Approaches to Integrating the MOVES Emissions Model with Traffic Simulators. Procedia Comput. Sci., The 4th International Conference on Ambient Systems, Networks and Technologies (ANT 2013), the 3rd International Conference on Sustainable Energy Information Technology (SEIT-2013) 19, 882–887. https://doi.org/10.1016/j.procs.2013.06.118

Zhong, H., Chen, K., Liu, C., Zhu, M., Ke, R., 2024. Models for predicting vehicle emissions: A comprehensive review. Sci. Total Environ. 923, 171324. https://doi.org/10.1016/j.scitotenv.2024.171324